# The HH 212 interstellar laboratory: astrochemistry as a tool to reveal protostellar disks on Solar System scales around a rising Sun

CLAUDIO CODELLA[1,2*], CECILIA CECCARELLI[2], CHIN-FEI LEE[3,4], ELEONORA BIANCHI[2], NADIA BALUCANI[5,2,1], LINDA PODIO[1], SYLVIE CABRIT[6,7], FREDERIC GUETH[8], ANTOINE GUSDORF[6,7], BERTRAND LEFLOCH[2], BENOIT TABONE[9,6]

[1] INAF, Osservatorio Astrofisico di Arcetri, Largo E. Fermi 5,50125 Firenze, Italy;
[2] Univ. Grenoble Alpes, Institut de Planétologie et d'Astrophysique de Grenoble (IPAG), 38401 Grenoble, France
[3] Academia Sinica Institute of Astronomy and Astrophysics, P.O. Box 23-141, Taipei 106, Taiwan
[4] Graduate Institute of Astronomy and Astrophysics, National Taiwan University, No. 1, Sec. 4, Roosevelt Road, Taipei 10617, Taiwan
[5] Dipartimento di Chimica, Biologia e Biotecnologie, Università degli Studi di Perugia, Perugia 06123, Italy
[6] LERMA, Observatoire de Paris, Ècole Normale Supérieure, PSL University, CNRS, Sorbonne Université, Paris, France
[7] Laboratoire de Physique de l'École Normale Supérieure, ENS, Université PSL, CNRS, Sorbonne Université, Université Paris-Diderot, Sorbonne Paris Cité, Paris, France
[8] IRAM, 300 rue de la Piscine, 38406 Saint-Martin-d'Hères, France
[9] Leiden Observatory, Leiden University, P.O. Box 9513, NL-2300 RA Leiden, the Netherlands

**ABSTRACT**

The investigation of star forming regions have enormously benefited from the recent advent of the ALMA interferometer working in the mm- and submm-wavelength spectral windows. More specifically, the unprecedented combination of high-sensitivity and high-angular resolution provided by ALMA allows one to shed light on the jet/disk systems associated with a Sun-like mass protostar. In this context, also astrochemistry enjoyed the possibility to analyze complex spectra obtained using large bandwidths: several interstellar Complex Organic Molecules (iCOMs; C-bearing species with at least 6 atoms) have been detected and imaged around protostars, often thanks to a large number of rotational-vibrational lines. This in turn boosted the study of the astrochemistry at work during the earliest phases of star formation paving the way to the chemical complexity in planetary systems where Life could emerge. There is mounting evidence that the observations of iCOMs (e.g. $CH_3CHO$ or $NH_2CHO$) can be used as unique tool to shed light, on Solar System scales (< 50 au), on the molecular content of protostellar disk. The increase of iCOMs abundances occur only under very selective physical conditions, such as those associated low-velocity shocks found where the infalling envelope is impacting the rotating accretion disk. The imaging of these regions with simpler molecules such as CO or CS is indeed paradoxically hampered by their high abundances and consequently high line opacities which do not allow the observers to disentangle all the emitting components at these small scales. In this respect, we review the state-of-the art of the ALMA analysis about the standard Sun-like star forming region in Orion named HH 212, associated with a pristine jet-disk protostellar system. We enrich the discussion with unpublished ALMA datasets, showing (i) how all the physical components involved in the formation of a Sun-like star can be revealed only by observing different molecular tracers, and (ii) how the observation of iCOMs emission, observed to infer the chemical composition of star forming regions, can be used also as unique tracer to image protostellar disks on Solar System scales, i.e. where planets will eventually form.

## 1. INTRODUCTION

A solar-type planetary system forms following a very complex process, which changes the diffuse gas and dust of a condensation inside a molecular cloud into a star surrounded by its planetary system. Simultaneously with this physical transformation, matter evolves chemically increasing its complexity at each further step (e.g. [1] for a review on the simultaneous physical and chemical evolution during the early phases of a solar-type star formation). Nowadays, evidence is mounting that the first steps of the process, namely when the protostar is in the so-called Class 0 and I source phases, are crucial for the future of the nascent planetary system, as discussed in the following.

Briefly, Class 0/I sources (see e.g. [2]) are the youngest ($10^4$-$10^5$ yr) solar-type protostars and are schematically composed by four major components: (1) a central stellar object, which will eventually become the star of the system (2) a circumstellar disk feeding the forming star by accretion, and where the planets, comets and asteroids of the system will eventually form: (3) an infalling envelope, which provides the matter that feeds the nascent planetary system; (4) a jet plus outflow system, which evacuates the angular momentum of the infalling material[1,3]. Class 0/I sources deliver their bolometric luminosity by accretion onto the stellar surface. Approximately, a star whose mass is 1 solar mass ($M_\odot$) after $10^5$ years of evolution, will have a bolometric luminosity of about 20 solar luminosities ($L_\odot$) (e.g. [4]). At the beginning, the envelope is massive enough to obscure the UV, optical and Near-Infra-Red (NIR) radiation emitted by the central object, so that only a cocoon of cold (about 10 K) material emitting in the sub-millimeter to millimeter (mm) wavelengths is visible. With time, the envelope will fade away exposing the warmer regions of the protostar and its disk, which becomes visible also at shorter wavelengths. To have an order of magnitude of the physical conditions, densities and temperatures of the envelopes go from about $10^4$ cm$^{-3}$ and 10 K in the outer zones to more than $10^8$ cm$^{-3}$ and 100 K in the innermost 100 au regions. The latter are called "hot-corinos" when they are also enriched of relatively complex organic molecules, which are C-bearing molecules with at least 6 atoms and that we will call in the following iCOMs for interstellar Complex Organic Molecules to make it clear that the term "complex" is only justified in the interstellar conditions[5,6].

As mentioned at the beginning, Class 0/I sources are becoming more and more important in the study of planet formation because evidence is mounting that the chemical composition of the final planetary system objects is highly influenced by the one during the Class 0/I phases. One illustrative example is represented by the terrestrial water, very likely, inherited from the very first phases of the Solar System life[7,8,9]. Another example is provided by the Solar System comets, whose ices have a chemical composition similar to that found in the hot-corinos[10,11,12].

In this context, it is of paramount importance to fully characterize the Class 0 and I phases, both from a physical and chemical point of view. The first difficulty in achieving this goal is that most of the actions, e.g. planets, comets and asteroids formation as well as the launching of the outflows and jets, occurs at a few tens astronomical units (au), which translates into the necessity of observations in the mm to Far-Infrared (FIR) with very high angular resolution (e.g. 0.1 arcsec for a source at 200 pc) in order to disentangles all the components (see above) and, consequently, the phenomena.

As we will discuss and show in this article, it turns out that chemistry is not only important per se to understand the link between the Solar System small bodies and its protostellar phases to shed light on its early history: chemistry is also a powerful tool to disentangle the various components at the tens au scale, allowing to study the different and complex processes which lead to the formation of a star plus its planetary system. Specifically, in the following, we will focus on this twofold role of the iCOMs. We will discuss in detail the illustrative case of HH 212, a prototype solar type Class 0 source with a well-known outflow jet, where numerous high spatial resolution observations have been obtained with the ALMA (Atacama Large Millimeter Array: https://www.almaobservatory.org) interferometer.

The article is organized as follows. Section 2 summarizes the set of ALMA observations used in this article, Section 3 describes the structure of HH 212 as derived by the observations of simple molecules (such as CO, CS and SiO), Section 4 discusses the powerful and unique role of iCOMs observations to reveal the thick disk surrounding the central future star, while Section 5 concludes this contribution.

## 2. THE ALMA DATASET OF HH 212

Table 1 lists all the lines that constitute the employed dataset here, with the spectroscopic properties, the relevant references and what component of the HH 212 system each transition best probes (see Sect. 3 and 4).

| Transition[a] | $\nu$[a] (GHz) | Eu[a] (K) | S$\mu^2$[a] (D$^2$) | Tracer | Ref. |
|---|---|---|---|---|---|
| SiO(8-7) | 347.3306 | 75 | 76.78 | Jet | a |
| $^{30}$SiO(8-7) | 338.9300 | 73 | 76.78 | Jet | b |
| C$^{17}$O(3-2) | 337.0611 | 32 | 0.04 | Cavity walls + Envelope | c |
| C$^{34}$S(7-6) | 337.3965 | 50 | 25.59 | Cavity walls + Envelope | c |
| HDCO(5$_{14}$-4$_{13}$) | 335.0968 | 56 | 26.05 | Cavity + Disk Atmosphere/Chemical Barrier | b |
| CH$_3$OH(2$_{2,1}$-3$_{1,2}$)A | 335.1337 | 45 | 0.31 | Disk Atmosphere/Chemical Barrier | b |
| CH$_3$OH(7$_{1,7}$-6$_{1,6}$)A | 335.5820 | 78 | 5.55 | Disk Atmosphere/Chemical Barrier | b |
| $^{13}$CH$_3$OH(12$_{1,11}$-12$_{0,12}$)A | 335.5602 | 193 | 22.97 | Disk Atmosphere/Chemical Barrier | d |
| CH$_3$CHO(18$_{0,18}$-17$_{0,17}$)A | 335.3587 | 155 | 226.88 | Disk Atmosphere/Chemical Barrier | e |

**Table 1:** List of molecular transitions used in the present article. Columns 1 to 4 report the spectroscopic properties of the transitions, Column 5 the component of the HH 212 system best probed by each line and the last column the reference to the published and new data. *References:* a. Lee et al. [13]; b. Present work; c. Tabone et al. [14]; d. Bianchi et al. [15]; e. Codella et al. [16]. *Note:*(a) Frequencies and spectroscopic parameters of $^{17}$CO, CH$_3$CHO are extracted from the Jet Propulsion Laboratory molecular database (JPL, Pickett et al. [17]), those of SiO, $^{34}$CS $^{30}$SiO, HDCO, CH$_3$OH and $^{13}$CH$_3$OH are from the Cologne Database for Molecular Spectroscopy (CDMS, Müller et al. [18,19]. Upper level energies refer to the ground state of each isomer.

Supporting Material reports: technical details of the data, and the whole channel maps of the CH$_3$OH(2$_{2,1}$-3$_{1,2}$)A and CH$_3$OH(7$_{1,7}$-6$_{1,6}$)A emissions.

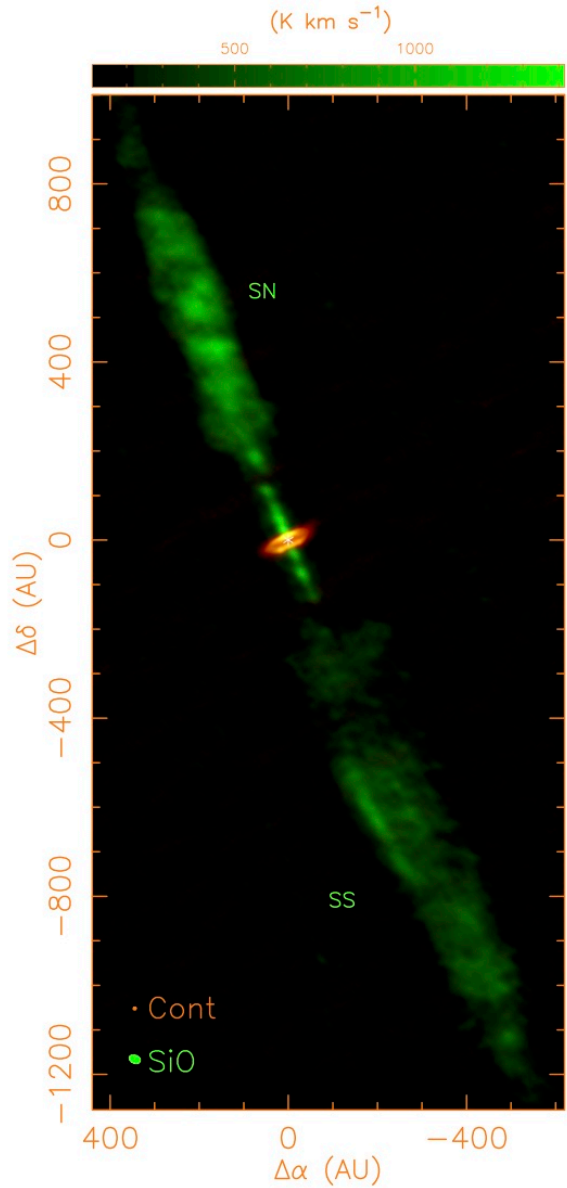

**Figure 1:** The image of the HH 212 protostar and its jet[13]. The green shows the emission from the SiO J=8-7, which very nicely traces the jet emanating from the central object, HH 212-mm, whose inner infalling-rotating envelope plus disk is seen using dust continuum emission at 352 GHz (orange image). The map is centered at $\alpha_{J2000} = 05^h 43^m 51.404^s$, $\delta_{J2000} = -01^o 02' 53.11''$.

# 3. HH 212: AN ILLUSTRATIVE CASE OF THE SOLAR TYPE STAR FORMATION PROCESS COMPLEXITY

The HH 212 star forming region, located at a distance 405(15) pc [20] in the L1630 cloud in Orion B, is one of the most spectacular and famous protostars and a perfect illustrative case to illustrate the complexity of a young protostar structure and of the processes occurring during the Class 0/I phases. The extended (240", 0.47 pc) bipolar symmetric jet which powers the outflow itself was first discovered by Zinnecker et al. [21] thanks to observations of the $H_2$ rotational-vibrational lines. Successively, all the four components of a Class 0/I protostar mentioned in the Introduction have been imaged[22,23,24,25]: a central object obscured by the envelope, a disk and an outflow plus jet system (see Figs. 1 and 2-Left). In the following, we will provide a review of what is known of these components, with emphasis on what molecules have mostly used to study each component.

## 3.1 The central object, its infalling-rotating envelope and its putative disk

The protostar, called HH 212-mm, is only detected in the (sub-)mm (e.g. [22]): it has a bolometric luminosity of 9 $L_\odot$ and a mass 0.25 $M_\odot$. Images of the infalling envelope have been obtained by ALMA[18] using lines from the $C^{17}O$, which is expected to be 1800 times less abundant than $C^{16}O$ (e.g. [26]). Its lower abundance allows to peer through the envelope and study the innermost regions, where the infalling-rotation velocity is higher and, consequently, easier to measure.

The middle panel of Figure 2 shows the spatial distribution of the $C^{17}O$ J=3-2 line emission in the innermost 400 au of the envelope[14]. The line emission has been binned over two velocity ranges, with respect to the systemic velocity (namely the velocity of the source with respect to the Sun): (i) the emission from the gas at less than 2 km s$^{-1}$, which shows the flattened shape of the envelope at these scales, and (ii) the gas at a negative/positive velocity between 2 and 4 km s$^{-1}$, which very nicely shows that the flattened envelope is rotating around the central object[25].

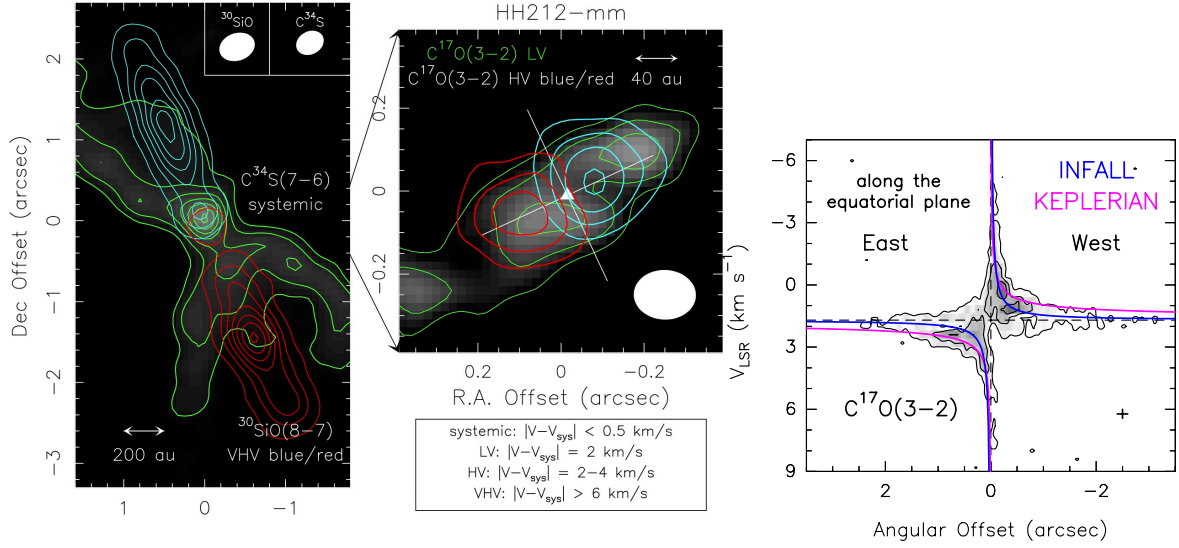

**Figure 2.** *Left Panel*: The HH 212 protostellar system (zoom-in of the large region shown in Fig. 1) as revealed by ALMA by combining different line tracers and different emission velocities. The maps are centered at $\alpha_{J2000}$ = 05$^h$ 43$^m$ 51.404$^s$, $\delta_{J2000}$ = -01$^o$ 02' 53.11". Blue/red contours plot the very high-velocity (VHV) blue/redshifted $^{30}$SiO(8-7) jet and the C$^{34}$S(7-6) asymmetric cavity at systemic velocity (grey background and green contours). Positions are with respect to the coordinates of the MM1 protostar, reported in Sect. 2. The C$^{34}$S map has been obtained by integrating the velocities from -0.8 km s$^{-1}$ to +0.8 km s$^{-1}$ with respect to the systemic velocity Vsys = +1.7 km s$^{-1}$ [24]: first contour and steps are 5$\sigma$ (10 mJy beam$^{-1}$ km s$^{-1}$) and 3$\sigma$, respectively. For $^{30}$SiO first contours and steps are 5$\sigma$ and 10$\sigma$ respectively: the blue map has been obtained by integrating down to -13 km s$^{-1}$ (1$\sigma$ = 5 mJy beam$^{-1}$ km s$^{-1}$), while the red map collects emission up to +11 km s$^{-1}$ (1$\sigma$ = 3mJy beam$^{-1}$ km s$^{-1}$). The filled ellipses show the synthetized beams (HPBW): 0.45" x 0.34" (-64$^o$), for $^{30}$SiO, and 0.44" x 0.35" (-63$^o$), for C$^{34}$S. *Middle Panel*: Zoom-in of the central region: C$^{17}$O(3-2) low-velocity (LV) emission, integrated over ±2 km s$^{-1}$ with respect to Vsys (grey background and green contours), overlaid to the blue/red high-velocity (HV), shifted by 2-4 km s$^{-1}$ with respect to Vsys. For C$^{17}$O first contours and steps are 5$\sigma$ (LV: 5 mJy beam$^{-1}$ km s$^{-1}$; HV red: 50 mJy beam$^{-1}$ km s$^{-1}$; HV blue: 65 mJy beam$^{-1}$ km s$^{-1}$) and 3$\sigma$, respectively. The ellipse shows the ALMA synthesized beam (HPBW): 0.14" x 0.13" (PA = -70$^o$). The white cross (oriented to illustrate the direction of the $^{30}$SiO(8-7) jet and consequently the equatorial plane) indicates the position of the protostar (white triangle). *Right panel:* position velocity (PV) cut of C$^{17}$O(3-2) perpendicular the jet direction (i.e. along the equatorial plane; see Figs. 1 and 2). Dashed lines mark the systemic velocity Vsys = +1.7 km s$^{-1}$ and the continuum peak associated with the protostellar disk[27]. First contour and steps are at 3$\sigma$ (22.8 K) and 5$\sigma$, respectively. The cross in the bottom-right corner is for the errors in both position and velocity. The keplerian curve R$^{-1/2}$ for a stellar mass of 0.25 M$_\odot$ (magenta) and the R$^{-1}$ curve for gas with angular momentum conservation (blue curve) are overplotted.

The map of the C$^{17}$O J=3-2 line emission can be used to provide more details on the gas envelope kinematics, specifically whether the gas moves following the Keplerian rotation of a rotationally supported disk. The right panel of Figure 2 shows this information can be extracted via the so-called PV (position-velocity) diagram[28]. At each position along the equatorial plane of the gas distribution (shown as the dashed line of Fig. 2-Right), the contours report the measured line intensity as a function of the observed velocity $v$ (i.e. the component projected along the line of sight), and of the distance R from the protostar. On the same figure, two theoretical cases are also shown: (i) rotationally supported disk in Keplerian motion around a 0.25 M$_\odot$ central object, i.e. $v(R)$ is proportional to $R^{-1/2}$, and (ii) rotating envelope with infall where the angular momentum is conserved. i.e. $v(R)$ is proportional to $R^{-1}$. In both cases, the velocity are increasing at low radius creating the so-called butterfly-pattern. The difference between the two cases is not large enough to distinguish which is correct in HH 212-mm, based on the observations of C$^{17}$O. In summary, in order to undoubtedly reveal the molecular component of the protostellar disk and to investigate its chemical composition one has to invoke different tracers with respect to tracers used for the protostellar environments (such as "simple" molecules, such as *CO and CS isotoplogues)*. We will see in Sect. 4 that iCOMs are a powerful diagnostic tool that can solves this ambiguity.

## 3.2 The outflow plus jet system

### 3.2.1 The jet

HH 212 has been the target of several observations using millimeter and sub-millimeter interferometers such as SMA, ALMA, and NOEMA[13,27,22,29,30,31,32,33], and whose goal has been to understand the dynamics and origin of the outflowing material. These observations have very often used simple abundant (CO) or shock-tracing (SiO) molecules, which revealed a fast (> 100 km s[-1]) jet with the following characteristics: beside some small-scale wiggling[13], the jet is associated with (i) a mirror symmetry either side of the HH 212-mm protostar[29,31], (ii) a stable direction on a scale of about 0.47 pc [21], and (iii) a spinning (i.e. rotation around its main axis) motion pointing to a launching radius of 0.05 au[24].

Figure 1 shows the image of the jet extending on each side of HH 212-mm[13]. The jet is probed by the SiO(8-7) line emission, a common tracer of shocks that suffers minimal contamination from the infalling envelope and other quiescent components of a protostar[13,31,34]. The reason is that, in cold molecular clouds and star forming regions, silicon is mostly trapped ($\geq$ 90%) in the dust silicate grains and in iced mantles (possibly containing Si-bearing species) enveloping them. When the grains are shattered by grain–grain collisions or the cores destroyed via gas–grain sputtering silicon and other Si-bearing species are injected into the gas, where Si quickly combines with oxygen and form additional SiO whose rotational lines emission is then detected[35,36,37]. In addition, since the SiO J=8-7 transition has a relatively large critical density ($10^8$ cm[-3]; the value after which the level population is dominated by collisions) it is particularly suited to probe dense gas. The reverse of the medal is that the line can easily be optically thick, preventing to probe the entire jet, as demonstrated by Cabrit et al. [32,33] in HH 212. For this reason, we obtained new observations of the rare isotopologue, $^{30}$SiO, with ALMA, whose technical details are described in the *Supporting Material*. Note that the $^{28}$Si/$^{30}$Si ratio has been recently confirmed to be ~30 in the protostar L1157 [38], which is similar to the value measured in the vicinity of the Sun[39]. The optical depth for $^{30}$SiO lines is then expected to be at least one magnitude less than those of the main isotopologues.

Figure 2-Left shows the ALMA [30]SiO image that we obtained: it is the first ever image of a protostellar jet traced by [30]SiO on 150 au scales. It reveals a pair of [30]SiO emission knots which emerge from HH 212-mm along a position angle of ~23º [40], in agreement with the previous maps of the jet[13,29,31]. What is most important in the context of this article, our new [30]SiO observations allow us to firmly establish the jet direction at small (≤150 au) scales and, consequently, identify the equatorial plane (normal to the jet main axis) where the accretion disk is expected to be placed, as we will discuss in Sect. 4.

### 3.2.2 The outflow cavity walls

The passage of the outflowing material creates cavities that are mostly emptied of gas and dust. Figure 2-Left shows the walls of such cavities in HH 212 as probed by the $C^{34}S(7-6)$ line emission around the systemic velocity[14]. This line has a relatively large critical density (~$10^9$ cm$^{-3}$) so that it traces the very dense gas, swept up by the passage of jet contributing to the wall of the cavity. Based on these images, the South cavities are clearly revealed, while only the East branch of the North cavity is visible, probably because of line excitation effects.

The new ALMA maps of the HDCO($5_{14}$-$4_{13}$) emission, shown in Figure 3 (Right panel), demonstrate that the cavity walls, observed at low velocities (< 1 km s$^{-1}$ with respect to the systemic velocity; see also [41]) are rotating in the same direction as the large-scale envelope (observed at ~$10^4$ au scales using $NH_3$ emission[42]) and the spinning jet at small scales (observed at ~1 au scale using SiO lines[14,24]. In addition, the HDCO images are suggestive that the cavity walls are made up of shocked material. This is because the deuterium fractionation of formaldehyde, as of any other molecule, is very low in warm (≥ 50 K) gas as it is very likely that of the cavity walls. Therefore, it is tempting to suggest that its detection in HH 212 points to the injection of HDCO from the grain mantles, which were formed in the prestellar cold phase and that are known to be highly deuterated[8]. To support this hypothesis, an enhancement of the $H_2CO$ deuterium fractionation has been previously detected in the cavity walls of another outflow system, L1157-B1 [43]. In other words, HDCO is a record of the very first phases of the HH 212 formation (see also [44]). Only future $H_2CO$ detection towards HH 212 will allow us to verify if deuteration is similarly enhanced in HH212.

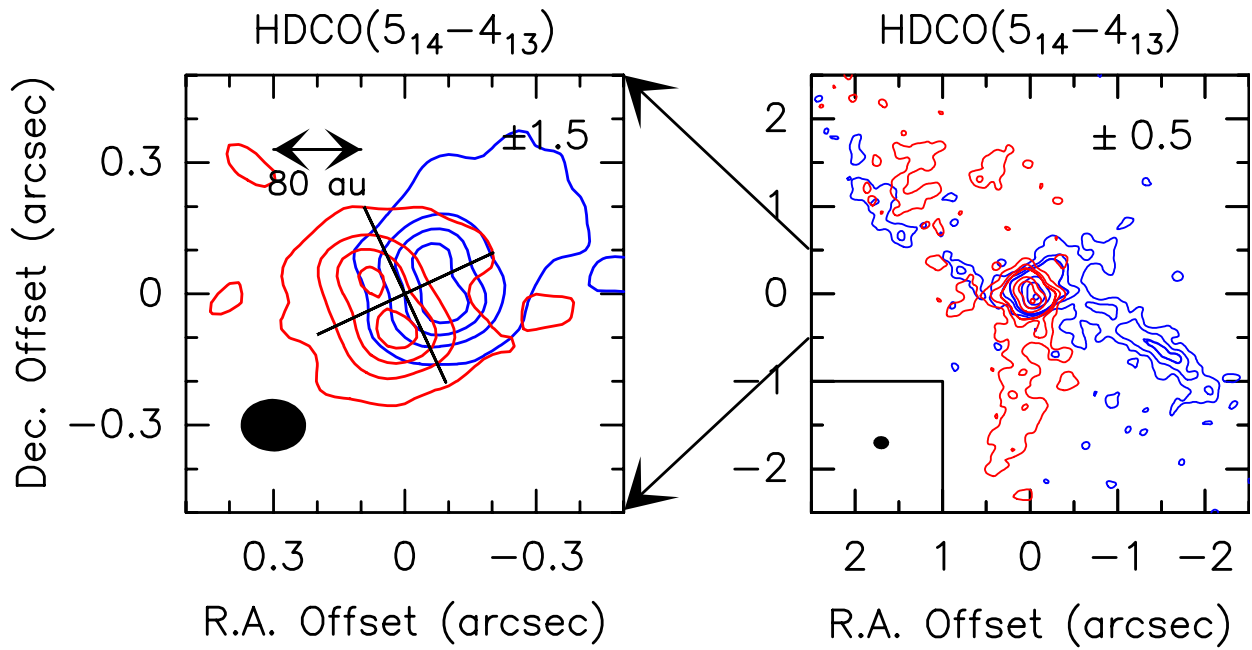

**Figure 3:** Channel maps of the HDCO($5_{1,4}$-$4_{1,3}$) blue- and redshifted emissions observed with ALMA towards the HH 212-mm protostar. Each panel shows the emission integrated over a velocity interval of 1.00 km s$^{-1}$ shifted with respect to the systemic velocity by the value given in the upper-right corner. The black cross (oriented to illustrate the direction of the SiO jet and consequently the equatorial plane, see Figs. 1 and 2) indicates the position of the protostar. The right panel shows a field of view 5" to image the largest scale structure at low velocities, cleaned weighting more the short *uv* baselines (see Appendix 1). In this case, the ALMA synthesized beam (HPBW) is 0.18" x 0.17" (PA = -72°). First contours correspond to 3σ (30.0 mJy beam$^{-1}$ km s$^{-1}$). Steps till 90 mJy beam$^{-1}$ km s$^{-1}$ correspond to 3σ to better show the extended cavities, then steps correspond to 20σ to avoid to graphically saturate the inner 1" region, which is shown in the left panel with a beam of 0.14" x 0.13" (PA = -88°) and steps of 10σ.

# 4. THE CENTRIFUGAL BARRIER PLUS DISK REVEALED BY THE iCOMs

In the previous section, we showed the incapacity of simple molecules like CO or CS to reveal the presence of a disk around HH 212, mostly because HH 212 is a very young system with a relatively massive infalling envelope, which might mask the possibly forming disk. If, on the one hand, the source youth is an obstacle from an observational point of view, on the other hand, it provides us with the (almost unique) possibility to study the very process of the formation of a disk, which will eventually evolve into a protoplanetary disk and then a planetary system. Following the pioneer work by Stahler et al. [4], a forming disk can be approximated by two main distinct regions: (i) the *outer disk*, which can be considered as the innermost part of the infalling-rotating envelope discussed in Sect. 3.1; and (ii) the *inner disk*, where the gas moves (quasi-)circularly following the Keplerian law. In addition, it is expected also a ring of shocked material[45,46] in front of the *centrifugal barrier*, which is the transition region between the inner and outer disk and where the excess energy and angular momentum are dispersed. In practice, the centrifugal barrier is the place where the infalling material cannot move inward without loosing energy (see Fig. 4 for a schematic illustration, adapted from Oya et al. [47]). In this section, we will focus on the inner disk and the centrifugal barrier, the two regions that simple molecules have difficulty to probe in HH 212 and that, as we will show here, can be revealed and studied via iCOMs.

The velocity of the accretion shock goes as $(M/Rc)^{0.5}$ (where M is the protostar mass and Rc is the centrifugal radius [48], and is typically of the order of 1 km s$^{-1}$. For this shock velocity, the post-shock temperature for the molecular gas is close to 100 K [48]. Which are the expected differences, in terms of chemistry, between accretion shocks and the shocks produced by the jet (see Sect. 3.2.1) moving through the protostellar environment? On the one hand, protostellar jets cause high-velocity (> 10 km s$^{-1}$) shocks cause sputtering and shattering leading to the erosion of the grain cores and ices, and consequently to the chemical enrichment of the gas phase [49,50]. On the other hand, accretion shocks are slower and they should not sputter or shatter the dust refractory core. As a consequence, the Si-bearing species are not expected to dramatically enhance their abundances in accretion shocks, conversely to what happens along the jet. However, the accretion shocks may still heat the grains enough to inject the material frozen on the dust mantles into the gas phase, and a drastic enrichment of the gas-phase material is the natural consequence.

The prediction of the existence of a ring where the angular momentum excess could be dispersed was first confirmed by Sakai et al. [45,46] towards the Class 0 source L1527. More detections of centrifugal barriers in other sources have followed thanks to the high sensitivity and spatial resolution of ALMA (e.g. [45,46,47,51] and references therein). In L1527, the centrifugal barrier was recognized thanks to the simultaneous observation of cyclopropenylidene (c-$C_3H_2$) and sulfur monoxide (SO); while the former is present in and probe the infalling-rotating envelope, SO shows an abrupt abundance increase in a ring of warm (~100 K) gas. The relatively large temperature and sudden SO abundance jump suggest the presence of a shock, interpreted by Sakai et al. [45,46] as the mild ($v_{shock}$ ~1 km/s) accretion shock in front of the centrifugal barrier. We notice that L1527 was known to be a source where iCOMs are absent and this explains why SO was used to trace the centrifugal barrier (see [52,53] for a discussion on the chemical diversity in Class 0 sources). On the contrary, towards HH 212, the SO emission is much more complex, as it is associated with the fast axial jet, with a slower rotating outflow emerging from the disk as well as the infalling-rotating envelope, so that it does not identify the centrifugal barrier[14,54,55].

Recently, Codella et al. [56] detected the presence of acetaldehyde in the small (40 au in radius) and dense ($\geq 10^9$ cm$^{-3}$) central region of HH 212, a clear signature that it harbors a hot-corino (see also [16,24,57,58]). This provides us with a precious tool, line emission from iCOMs, (i) to probe the inner 40 au region, (ii) to investigate possible effects due to the optically thick continuum emission along the equatorial plane, and, (iii) to study the HH 212 centrifugal barrier and inner disk. In the following, we report the information extracted from ALMA observations of acetaldehyde and methanol.

## 4.1 Acetaldehyde: a rotating ring of warm gas

ALMA high spatial resolution observations to the acetaldehyde emission in the central innermost HH 212 region has been obtained by Codella et al. [16]. Figure 5 shows the acetaldehyde ($CH_3CHO$) emission map and its PV (position-velocity) diagram (see Sect. 3.1 for the explanation of this diagram). More specifically, we show the emission at systemic velocity (Fig. 5-Middle) as well as the emission blue-/redshifted by about 2 km s$^{-1}$ (Fig. 5-Left) (see [16] for the whole channel maps).

First, the emission is compact and confined in a region whose radius is about 40 au. In addition, the analysis of the several $CH_3CHO$ detected lines showed that the gas has a temperature larger than about 80 K. This temperature is consistent with both a shocked and a thermally heated (from the central star) gas origin (see [16]). In the first case, the jump in the acetaldehyde abundance would be caused by the shock sputtering of the iced grain mantles (a process similar to that discussed in Sect. 3 for HDCO and Si-bearing species trapped in the mantle); in the second case, it would be due to the sublimation of the water-rich grain mantles. In both cases, acetaldehyde would be either just extracted by the icy mantles or formed in the gas phase from freshly injected mother species previous trapped in the mantles.

The second important information provided by the acetaldehyde map is that, contrarily to the PV diagram of $C^{17}O(3-2)$ (shown in Sect. 3.1), the $CH_3CHO$-A($18_{0,18}-17_{0,17}$) PV clearly demonstrates that this line is emitted in a ring of gas, which rotates around the center as a rigid body. A rotating ring shows its radial (i.e. projected along the light of sight) maximum and minimum velocities at the maximum distance (projected on the plane of the sky) from the protostar, while the velocity is null towards the protostar itself. More precisely, $v(R)$ is proportional to $R$, as well shown by the $CH_3CHO$ PV of Fig. 5. Note that, as reported in Sect. 3.1, an extended Keplerian disk would be characterized by a different trend: $v(R) \sim R^{-1/2}$. In other word, acetaldehyde is consistent with tracing the shock in front of the centrifugal barrier.

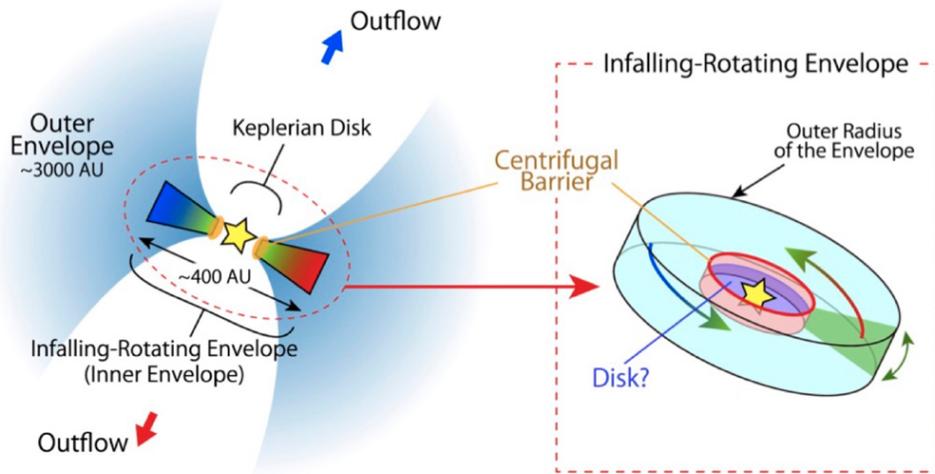

**Figure 4.** Schematic illustration [47] of the interface between the infalling envelope and the rotating disk.

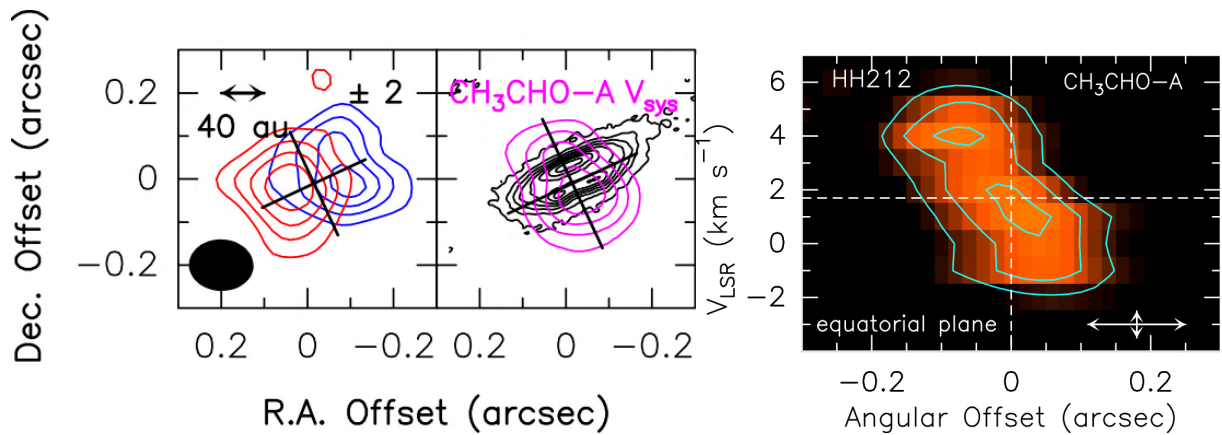

**Figure 5.** Adapted from Codella et al. [16]: *Left and middle panels*: Channel maps of CH$_3$CHO($18_{0,18}-17_{0,17}$) A blue- and redshifted emissions (by 2 km s$^{-1}$) observed towards the HH 212-mm protostar. The left panel shows the emission integrated over a velocity interval of 1 km s$^{-1}$ shifted with respect to the systemic velocity (see the magenta channel) by the value given in the upper-right corner. The emission at systemic velocity is overlaid on top of the disk traced by [23] using ALMA-Band 7 continuum observations (black contours). The black cross (oriented to illustrate the direction of the CO jet and consequently the equatorial plane, see Fig. 1) indicates the position of the protostar. The ellipse in the lower left panel shows the ALMA synthesized beam (HPBW): 0.14" x 0.12" (PA = 84°). First contours and steps correspond to 3σ (3.0 mJy beam$^{-1}$ km s$^{-1}$). *Right panel*: Position-velocity cut of CH$_3$CHO($18_{0,18}-17_{0,17}$) A emission (colour scale, cyano contours) along the equatorial plane. First contours and steps correspond to 3σ (3.1 K km s$^{-1}$). Dashed lines mark the position of HH 212-mm protostar and the cloud Vsys = +1.7 km s$^{-1}$; [27]). The error bars are drawn In the bottom-right corner.

## 4.2 The hamburger shape of methanol emission: disk atmosphere or accretion shock?

In order to understand better the structure of the ring and, consequently, the centrifugal barrier, the (brighter) emission from another iCOM, methanol, turns out to be instrumental. The PV of the methanol emission, as shown by Lee et al. 24, is dominated by the same linear pattern as CH$_3$CHO (shown in Fig. 5), We can thus argue that their emissions are both dominated by a bright rotating ring at about the same radius. The brightness of the methanol lines allows us to infer more information on the properties of the ring of iCOMs emission, revealed by CH$_3$CHO. Figure 6 shows the shape of the CH$_3$OH emission at the systemic velocity (namely the gas with no movement with respect to the center of the system) as well as that at a negative/positive velocity of about 1 km s$^{-1}$ (which shows the gas rotation; note that all the channel maps are reported in Fig. S.1). These maps reveal that the CH$_3$OH emission is not confined towards the equatorial plane of the disk but, on the contrary, peaks above and below it, i.e. towards the South and North poles (see also [24,58]).

This is even better seen in Figure 7, exploiting the $CH_3OH$ data obtained at different spatial resolutions. We show a spatial resolution enhanced methanol image overlapped with the millimeter continuum emission, which probes the dusty disk. More specifically, the upper panel of Fig. 7 shows the red- and blue-shifted emission in stacked maps of $CH_3OH$ lines with Eu in the 44-332 K range, observed with a spatial resolution of 12 au, on top of the dusty disk continuum map[24]. On the other hand, Fig. 7-Bottom reports the distribution of $CH_3OH(7_{1,7}-6_{1,6})$A centroid positions in the velocity channels (0.44 km s$^{-1}$ wide). In this case, the spatial resolution is 60 au, but the Signal-to-Noise (S/N) is high enough to allow us to fit models directly through the so-called *uv* interferometic visibilities. The error on centroid position provided by *uv*-fit is the function of the channel S/N and atmospheric seeing, and is typically much smaller than the beam size. For instance, an un-limited over-resolution power can, in principle, be achieved if the dynamic range of the observations is arbitrarily large[59]. In other words, using bright lines it is possible to infer the spatial distributions at scales definitely smaller that the nominal HPBW beam. In this case, the error on the centroid position is less than 20 mas, i.e. less than 8 au. The fit has been performed assuming two gas components associated with the northern and southern outer disk. Magenta points are for the emission in the channel sampling the systemic velocity, while red and blue points are for the channels red- and blue-shifted in the velocity range (with respect to Vsys = +1.7 km s$^{-1}$; [27]) identified by the labels.

Considering that the extension across the disk seen in $CH_3OH$ channel maps and centroids is due to projection effects along the ring revealed by the PV, from the analysis of the images of Fig. 7, methanol seems to trace the rotating buns of a hamburger with the dusty disk in the role of the meat. At high velocities, we clearly spatially resolve the red- and blue-shifted regions that move away from the jet axis, revealing the velocity gradient. In principle, all these findings have two explanations: (1) the disk has a vertically extended gaseous atmosphere (the "methanol buns") and no gaseous methanol on the disk plane (the "dusty meat"); (2) the "dusty meat" disk is optically thick and obscures the methanol emission behind it so that only the disk atmosphere $CH_3OH$ emission can escape. However, in order to firmly conclude which of the two above explanations is the correct one, high spatial resolution observations are needed at much lower frequencies, e.g. in the radio, where the dust continuum is much more likely optically thin.

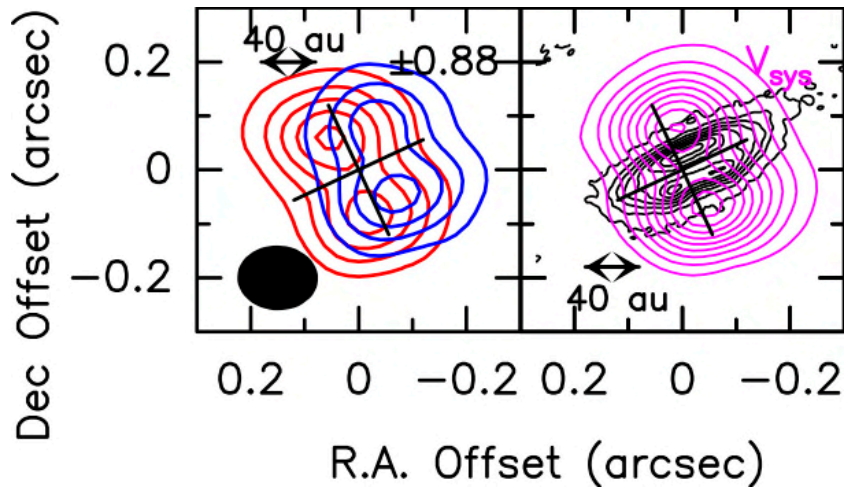

**Figure 6.** Representative channel maps of the $CH_3OH(7_{1,7}-6_{1,6})E$ blue- and redshifted emissions observed with ALMA towards the HH 212-mm protostar. *The whole channel maps, including those of $CH_3OH(7_{1,7}-6_{1,6})A$, are reported in Appendix 2.* The left panel shows the emission integrated over a velocity interval of 0.44 km s$^{-1}$ shifted with respect to the systemic velocity (see the right channel) by the value given in the upper-right corner. The black cross (oriented to illustrate the direction of the CO jet and consequently the equatorial plane, see Fig. 1) indicates the position of the protostar. The ellipse in the lower left panel shows the ALMA synthesized beam (HPBW): 0.14" x 0.12" (PA = 84°). First contours and steps correspond to 5σ (10.0 mJy beam$^{-1}$ km s$^{-1}$) and 5σ, respectively.

Having said that, there might be a gradient in the gas temperature moving higher in the atmosphere. The excitation temperature of the methanol emission as imaged by Lee et al. [24] above and below the equatorial plane is 150±50 K, consistent with what found for acetaldehyde (larger than 80 K). Indeed, only the detection of sublimated species on the midplane of the disk could test this hypothesis. The edge-on geometry, which is a plus for the detection of the external disk layers, is paradoxically hampering a search for the molecular gas along the equatorial plane. However, this problem can be overcome by moving from (sub-)mm to cm-wavelengths (using e.g. the NRAO VLA interferometer), where continuum emission is expected to be definitely thinner. Should a vertical temperature gradient be confirmed (higher in the outer layers), the hypothesis that methanol and acetaldehyde probe the shock in front of the centrifugal barrier may be weakened in favor of the hypothesis that both species are (simply) sublimated from icy dust mantles and warmed up by the central source, because there is no (obvious) reason of a vertical temperature gradient in a shock at the centrifugal barrier. Another possibility is that the apparent disk atmosphere is part of the wind observed in SO emanating from the disk and at the base of the jet [14,16,54]. Again, more high-spatial resolution observations are necessary to fully understand the structure of the centrifugal barrier and inner disk in HH 212 and the contribution of the disk wind.

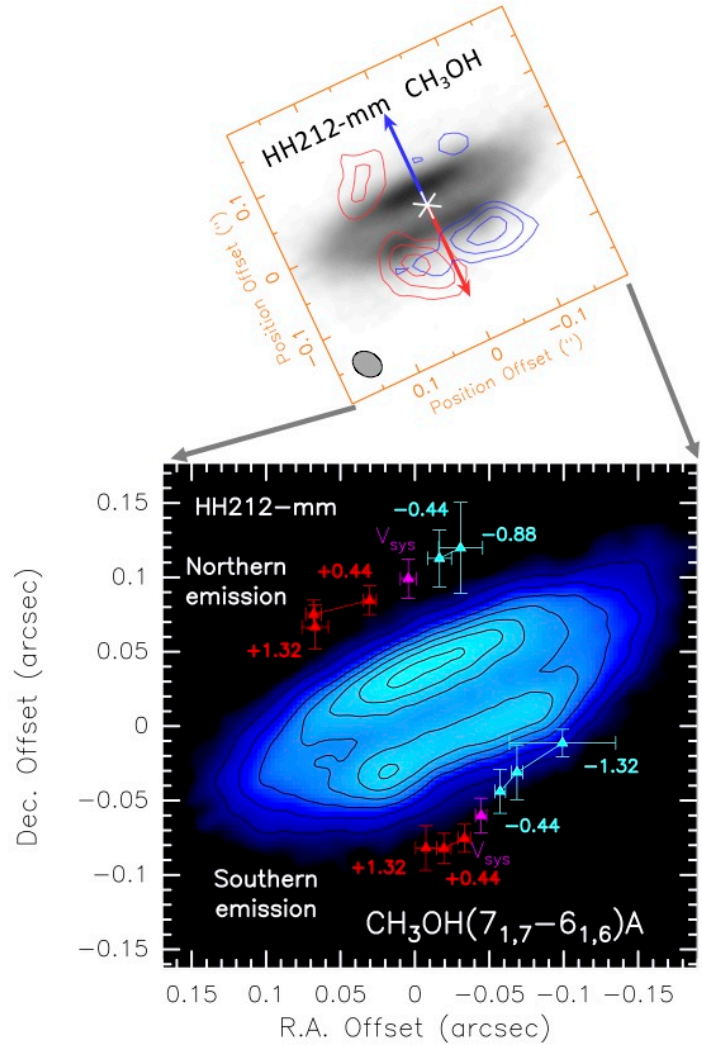

**Figure 7.** *Upper panel:* Red- and blue-shifted emission in the stacked maps of $CH_3OH$ lines with Eu in the 44-332 K range on top of the dusty disk continuum map[24]. *Bottom panel:* Distribution of $CH_3OH(7_{1,7}-6_{1,6})$A centroid positions (from fits in the *uv* domain with 1σ error bars) in the velocity channels (sampled at the original spectral resolution of 0.44 km s$^{-1}$). The fit has been performed assuming two gas components (following [58], see text), associated with the northern and southern outer disk. Magenta points are for the emission in the channel sampling the systemic velocity, while red and blue data points denote the channels red- and blue-shifted in the velocity range (with respect to Vsys = +1.7 km s$^{-1}$; [27]) identified by the labels. The points are overlaid on top of the disk traced by [23] using ALMA-Band 7 continuum observations (color scale and contours).

**4.3 Acetaldehyde and methanol abundance ratios: feeding astrochemical models**

Acetaldehyde and methanol are not the unique iCOMs so far detected towards the HH 212 40 au inner zone: formamide ($NH_2CHO$), methanethiol ($CH_3SH$), formic acid ($HCOOH$), methyl formate ($HCOOCH_3$), acetonitrile ($CH_3CN$) and ethanol ($CH_3CH_2OH$) have been very recently added to the list[24,58]. Thanks to the use of astrochemical models, the spatial distribution of the measured abundance ratio of iCOMs can lead to very stringent constraints on how they are formed and destroyed[49]. In particular, they showed the formamide over acetaldehyde abundance ratio towards the L1157-B1 shock demonstrating a gas-phase formation route of formamide[60]. It is, hence, worth to analyze in some detail the abundance ratio of the two iCOMs in HH 212 with the brightest lines: acetaldehyde and methanol.

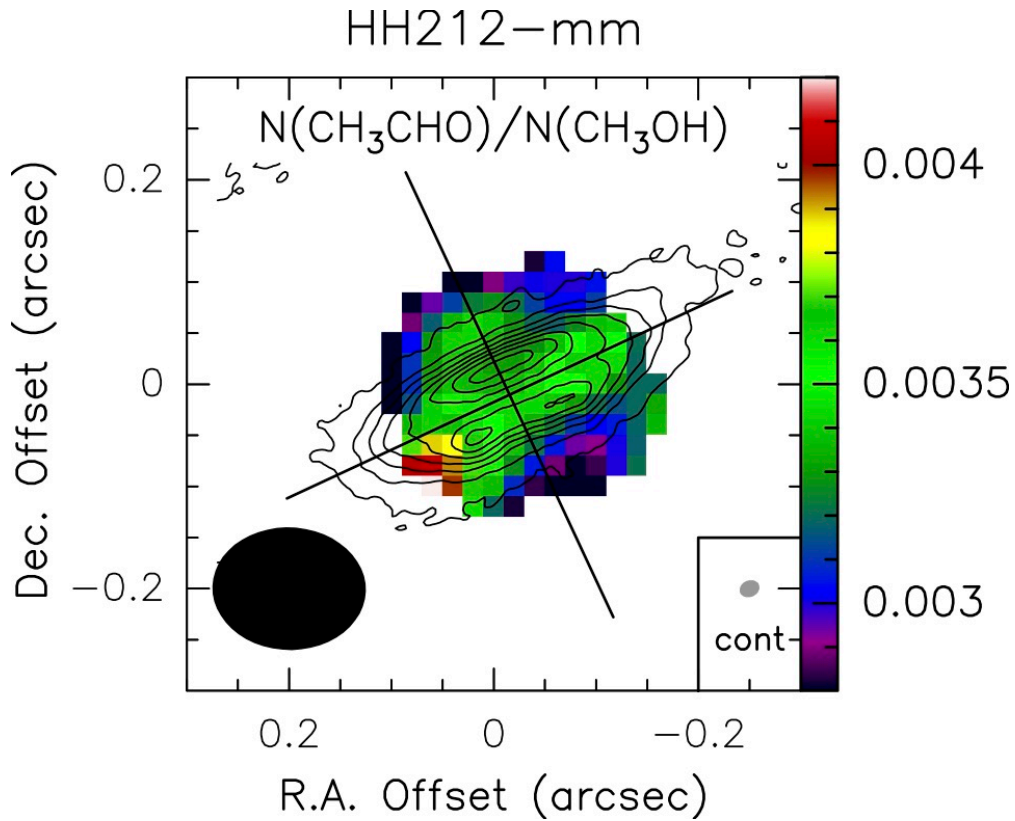

**Figure 8.** Map in color scale of the ratios (where both emissions are at least 3σ) between the column densities of $CH_3CHO$ [16] and $^{13}CH_3OH$ (multiplied by 50 to take into account the $^{12}C/^{13}C$ isotopic ratio as measured in Orion; [61]), overlaid on top of the disk traced by Lee et al. [23] using ALMA Band 7 continuum

observations (black contours). The black cross (inclined in order to point the CO jet direction and consequently the equatorial plane, see Fig. 1) indicates the position of the protostar. The ellipse in the bottom left (right) shows the ALMA synthesized beam (HPBW) for line (continuum) images.

To this end, we estimated the $CH_3CHO$ abundance with respect to $CH_3OH$, which can be safely assumed as formed on dust mantles and then injected into the gas phase. Assuming that both species are populating the same gas portion (as demonstrated by the same PV pattern, and confirmed by recent acetaldehyde images obtained stacking several lines [58]), the ratio between their column densities can be used to derive their abundance ratio. In order (i) to minimize line opacity effects and (ii) avoid calibrations effects we derived the methanol column density starting from the $^{13}CH_3OH$ emission (*observed with the same ALMA Cycle 4 dataset and an angular resolution of 0.15" (60 au) used to image CH₃CHO*). The column densities have been derived assuming Local Thermodynamic Equilibrium (LTE) conditions, supported by the high densities found ($10^9$ $cm^{-3}$). The $^{13}CH_3OH$ emission can be assumed optically thin, given the opacity of the main isotopologue has been found to be less than 0.4 [57]. We then assumed a $^{12}C/^{13}C$ ratio of 50 following the results on the Orion complex (where HH212 is located) reported by Kahane et al. [56], and adopted by Lee et al. [58].

We used the rotational diagram analysis to determine the temperature and the column density. For a given molecule, the relative population distribution of all the energy levels, is described by a Boltzmann temperature, that is the rotational temperature Trot. The upper level column density can be written as:

$$N_u = (8 \pi k \nu^2) \times (h c^3 A_{ul})^{-1} \times (\eta_{bf})^{-1} \times \int T_b \, dV$$

where k is the Boltzmann constant, $\nu$ is the frequency of the transition, h is the Plank constant, c is the light speed, $A_{ul}$ is the Einstein coefficient, $\eta_{bf}$ is the beam-filling factor and the integral is the integrated line intensities (in brightness temperature scale). Nu is related to the rotational temperature Trot, as follow:

$$\ln(Nu/gu) = \ln(Ntot) - \ln(Q(Trot)) - (Eu/k\ Trot)$$

where gu is the generacy of the upper level, Ntot is the total column density of the molecule, Q(Trot) is the partition function at the rotational temperature and Eu is the energy of the upper level.

Figure 8 shows the $CH_3CHO$ over $CH_3OH$ abundance ratio, overlaid on top of the disk dust continuum[16]. The ratio looks pretty constant, lying in the 3-4 x $10^{-3}$ range. The average acetaldehyde column density is N($CH_3CHO$) = 8(3) x $10^{15}$ cm$^{-2}$, while the $^{13}CH_3OH$ column density is 7(2) x $10^{16}$ cm$^{-2}$, which in turn leads to N($CH_3OH$) = 4(1) x $10^{18}$ cm$^{-2}$. Note that the $CH_3CHO$ and $^{13}CH_3OH$ column densities are consistent, within a factor 2-3, with those derived on smaller (16 au) spatial scales[24,58]. The very small variation of the $CH_3CHO/CH_3OH$ abundance ratio after $10^4$ years (the typical age of a protostar) suggests that the chemistry of these two species is plausibly dominated by the sublimation of the species and reactions in the gas phase. In other words, the fact that material may fall into the inner region of HH 212 does not have any consequence on the chemical composition of the gas, probably because the reactions are extremely fast. This is plausibly due to the large density of the gas, $10^8$-$10^9$ cm$^{-3}$. Only future imaging of more iCOMs as well as of the infalling gas will allow us to fully investigate this speculation.

It is interesting to check whether the $CH_3CHO/CH_3OH$ abundance ratio derived for the HH 212 disk is different with respect to what measured towards classical Class 0 and Class I hot-corinos. If we select only the sources where the inner 100 au around the protostars has been sampled using interferometers, the number is limited (see Table 2): IRAS16293-2422B [62,63], SVS13-A [12,64], B1b-S [65], L483 [66], and NGC1333-IRAS4A2 [67,68].

| Targets | Object Type | $CH_3CHO/CH_3OH$ | Ref. |
|---------|-------------|------------------|------|
| HH 212-mm | Class 0 | $1\text{-}3 \times 10^{-3}$ | a,b,c,d |
| NGC1333-IRAS4A2 | Class 0 | $3\text{-}6 \times 10^{-3}$ | e,f |
| IRAS16293-2422B | Class 0 | $10^{-2}$ | g,h |
| Barnard 1b-S | Class 0 | $2 \times 10^{-3}$ | i |
| L483 | Class 0 | $5 \times 10^{-3}$ | j |
| SVS13-A | Class I | $10^{-3}$ | k,l |
| 67P-CG | Comet | $3\text{-}4 \times 10^{-2}$ | m |
| Lovejoy | Comet | $5 \times 10^{-2}$ | n |
| Hale-Bopp | Comet | $4 \times 10^{-2}$ | o |

**Table 2:** Comparison between $CH_3CHO$ and $CH_3OH$ abundance ratios as sampled towards protostars on spatial scales less than 100 au. References: a. Present paper; b. Lee et al. 24; c. Codella et al. 16; d. Bianchi et al. 15; e. López-Sepulcre et al. 67; f. Taquet et al. 68; g. Jørgensen et al. 62; h. Jørgensen et al. 63; i. Marcelino et al. 65; j Jacobsen et al. 66; k. De Simone et al. 64; l. Bianchi et al. 12; m. Le Roy et al. 69; n. Biver et al. 70; o. Bockelée-Morvan et al. 71.

The values as measured towards Class 0 and I objects fall in the $10^{-3}$ - $10^{-2}$ range, indicating that the $CH_3CHO/CH_3OH$ emission from the HH 212 disk outer layers is similar to what measured towards classical hot-corinos. The comparison poses new questions: which is the spatial distribution of the chemically enriched gas observed towards the classical hot corinos? Does it mainly come from the outer layers of the protostellar disk? What process (shock sputtering versus thermal sublimation of the ices) is then responsible for injecting/forming the iCOMs in these objects?

Table 2 also lists what measured towards comets[69,70,71], where the $CH_3CHO/CH_3OH$ abundance ratio looks to be higher than fews $10^{-2}$. In other words, the aceltaldehyde abundance with respect to dust grain products such as methanol seems to increase from protostars to comets, which likely constitutes the most primitive material in the Solar System. Although this trend has to be verified increasing the numbers of observations of both evolutionary stages, it suggests some processing during the formation of a planetary system, questioning the inheritance from the protostellar phase. An alternative possibility is that different ratios could reflect different sublimation processes active in protostars and comets, the latter possibly associated with a thermal desorption.

To conclude, the analysis based on $CH_3CHO$ and $CH_3OH$ indicates that further investigations at high-spatial resolution are needed: (i) to observe more iCOMs towards HH 212-mm in order to better characterize the chemistry of its protostellar disk, (ii) to verify whether also classical hot-corinos are associated with chemical rich outer layers of a dusty disk, and (iii) to perform a reliable comparison with the measurements with comets.

## 5. SUMMARY

We reviewed the results recently obtained on the HH 212-mm protostar using the ALMA interferometer, which allowed us to shed light on Solar System scales on the chemistry occurring around a Sun-like forming star. Where the most abundant molecules fail in revealing the gas associated with the accretion disk, mainly due to line opacities, the emission due to roto-vibrational transitions of iCOMs has been successful. The reason resides in the selective physical conditions which allow iCOMs to enhance their abundance either due to a direct release into the gas-phase from the dust mantles or to a formation directly in the gas-phase using simpler species injected from grains. The physical processes which can enrich the gas-phase close to the protostars are numerous, such as thermal evaporation or sputtering. Whatever the mechanism is, the HH 212 case shows a chemical enrichment only above and below the dusty disk, within a radius of about 40 au. Note that so far HH 212 is the unique source where it is possible to compare the spatial distribution of the sub-mm disk dust with that of the iCOMs emission. Table 1 summarizes, for several physical components at work in the star forming process, which is the best tracer to use, according to the HH 212 mm surveys. The reasons for the enhancement of the iCOMs abundances have to be still clearly understood and they are indeed hotly debated. A rotating ring at a radius of the centrifugal barrier (40 au) is surely present. A possibility could be the occurrence of low-velocity shocks located where the infalling envelope is impacting the rotating accretion disk. An alternative possibility is that iCOMs are not associated with accretion shocks, but they are instead tracing the irradiated disk atmosphere, i.e. the outermost layers of the protostellar disk. Indeed, 40 au is also the radius where the thermal heating due to the protostellar radiation is expected to produce a dust temperature around 100 K [72], high enough to induce the evaporation of the dust mantles. If so, an open question is if iCOMs are linked with the dusty disk, or, conversely, they are leaving the system possibly contributing to form a wind from the disk.

Finally, we discussed the HH 212-mm chemical richness, which looks similar to classical protostellar hot-corinos, so far observed at lower spatial resolutions with respect to the HH 212 observations. Further investigations are needed: (i) to observe more iCOMs towards HH 212-mm in order to better characterize the chemistry of its protostellar disk, and (ii) to verify whether also classical hot-corinos are associated with chemical rich outer disk layers.

AUTHOR INFORMATION

**Corresponding Author**

Claudio Codella

codella@arcetri.astro.it

**Author Contributions**

The manuscript was written through contributions of all authors. All authors have given approval to the final version of the manuscript.

**Supporting Information**

Technical details of the data; whole ALMA channel maps of the $CH_3OH(2_{2,1}-3_{1,2})A$ and $CH_3OH(7_{1,7}-6_{1,6})A$ emissions.

ACKNOWLEDGMENT

This paper makes use of the ADS/JAO.ALMA#2016.1.01475.S data (PI: C. Codella). ALMA is a partnership of ESO (representing its member states), NSF (USA) and NINS (Japan), together with NRC (Canada) and NSC and ASIAA (Taiwan), in cooperation with the Republic of Chile. The Joint ALMA Observatory is operated by ESO, AUI/NRAO and NAOJ. This work was supported by (i) the PRIN-INAF 2016 "The Cradle of Life - GENESIS-SKA (General Conditions in Early Planetary Systems for the rise of life with SKA)", (ii) the program PRIN-MIUR 2015 STARS in the CAOS - Simulation Tools for Astrochemical Reactivity and Spectroscopy in the Cyberinfrastructure for Astrochemical Organic Species (2015F59J3R, MIUR Ministero dell'Istruzione, dell'Università della Ricerca e della Scuola Normale Superiore), (iii) the European Research Council (ERC) under the European Union's Horizon 2020 research and innovation programme, for the Project "The Dawn of Organic Chemistry" (DOC), grant agreement No 741002, and (iv) the European MARIE SKŁODOWSKA-CURIE ACTIONS under the European Union's Horizon 2020 research and innovation programme, for the Project "Astro-Chemistry Origins" (ACO), Grant No 811312. C.-F.L. acknowledges grants from the Ministry of Science and Technology of Taiwan (MoST 107-2119-M-001-040-MY3) and the Academia Sinica (Career Development Award and Investigator Award).

**SUPPORTING MATERIAL**

*Appendix 1: Technical details of the new ALMA observations presented in this review*

The new observations of HH 212 analyzed here were obtained with ALMA Band 7 using 34 12-m antennas between 15 June and 19 July 2014 during the Cycle 1 phase ($^{30}$SiO) and using 44 12-m antennas between 6 October and 26 November 2016 during the Cycle 4 phase (HDCO, $CH_3OH$, and $CH_3OD$). The maximum baselines for Cycle 1 and 4 were 650 m and 3 km, respectively. In Cycle 1 the spectral windows between 337.1-338.9 GHz and 348.4–350.7 GHz were observed using spectral units of 488 kHz (0.42–0.43 km s$^{-1}$). Calibration was carried out following standard procedures, using quasars J0607-0834, J0541-0541, J0423-013, J0510+1800, J0552+0313, J0541-0211 and J0552-3627, as well as Ganymede. The continuum-subtracted Cycle 1 images have clean-beam FWHMs from 0.41" x 0.33" to 0.52" x 0.34" (PA = -63º), i.e. about 160 au, and an rms noise level of ~30 mJy beam$^{-1}$. For Cycle 4 spectral units of 122 kHz (0.1 km s$^{-1}$) were used to observe the spectral windows between 335.0–337.4 GHz. The continuum-subtracted images have a typical clean-beam FWHM of 0.15" x 0.12" (PA = -88º), i.e. about 60 au, and an rms noise level of 4–5 mJy beam$^{-1}$. The CASA package was used to obtain spectral line imaging, while data analysis was performed using GILDAS (http://www.iram.fr/IRAMFR/GILDAS). The center of the maps here reported are given with respect to the MM1 protostar continuum peak located at R.A.(J2000) = 05$^h$ 43$^m$ 51.41$^s$, Dec(J2000) = -01º 02' 53.17" [27]. All the emission lines were identified using the Jet Propulsion Laboratory (JPL; spec.jpl.nasa.gov/, [17]) and Cologne Database for Molecular Spectroscopy (CDMS; https://cdms.astro.uni-koeln.de/; [18,19]) molecular databases.

In addition, this article uses ALMA (Band 7 - Cycle 4) observations of $C^{17}O$, $C^{34}S$, $^{13}CH_3OH$, and $CH_3CHO$ data, which are published in Tabone et al. [14], Codella et al. [16], and Bianchi et al. [15]. Table 1 of the Main Body summarizes the transitions used in the present article.

### Appendix 2: Additional maps of the methanol emission in the central 100 au

Figure S.1 reports the channel maps of the $CH_3OH(2_{2,1}-3_{1,2})A$ and $CH_3OH(7_{1,7}-6_{1,6})A$ emissions observed with ALMA (during Cycle 4) towards the HH 212-mm protostar. Each panel shows the emission integrated over a velocity interval of 0.44 km s$^{-1}$ blue- and redshifted with respect to the systemic velocity. The black cross is oriented to illustrate the direction of the CO jet and consequently the equatorial plane. The maps are discussed in Sect. 4.2 of the Main Body.

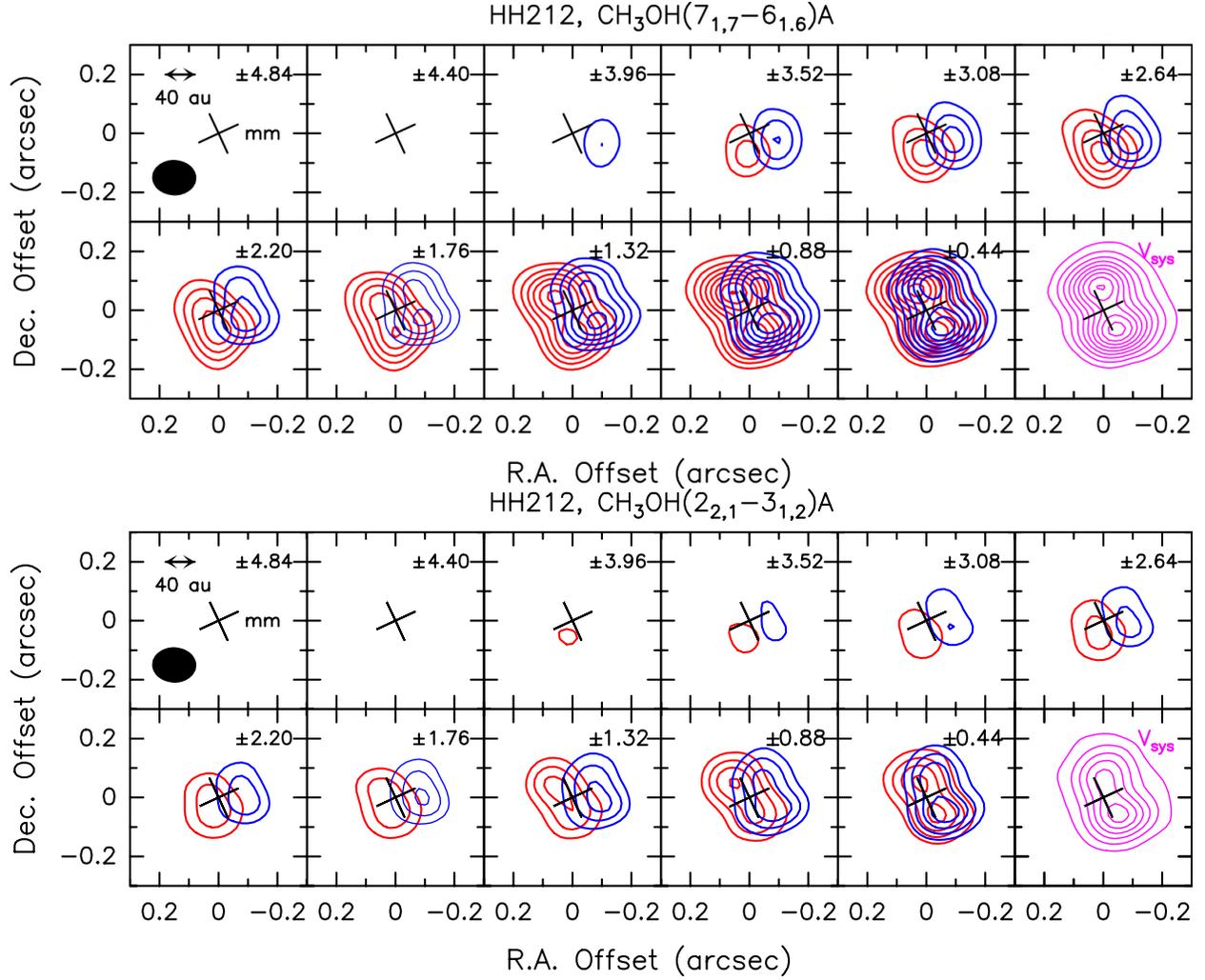

**Figure S.1.** Channel maps of the CH$_3$OH($7_{1,7}$-$6_{1,6}$)A (Upper panels) and CH$_3$OH($2_{2,1}$-$3_{1,2}$)A (Lower panels) blue- and redshifted emissions observed with ALMA towards the HH 212-mm protostar. Each panel shows the emission integrated over a velocity interval of 0.44 km s$^{-1}$ shifted with respect to the systemic velocity (see the magenta channel) by the value given in the upper-right corner. The black cross (oriented to illustrate the direction of the CO jet and consequently the equatorial plane, see Fig. 1) indicates the position of the protostar. The ellipse in the top left panel shows the ALMA synthesized beam (HPBW): 0.14" x 0.12" (PA = 84°). First contours and steps correspond to 5σ (10.0 mJy beam$^{-1}$ km s$^{-1}$) and 3σ, respectively.